\documentclass[preprint]{aastex62}

\accepted{23 December 2021}
\submitjournal{AJ}

\shorttitle{RSG mass loss}
\shortauthors{Humphreys and Jones}

\begin{document}

\title{Episodic Gaseous Outflows and Mass Loss from Red Supergiants }

\correspondingauthor{Roberta Humphreys}
\email{roberta@umn.edu}

\author{Roberta M. Humphreys}
\affiliation{Minnesota Institute for Astrophysics,  
University of Minnesota,
Minneapolis, MN 55455, USA}

\author{Terry J. Jones}
\affiliation{Minnesota Institute for Astrophysics,
University of Minnesota,
Minneapolis, MN 55455, USA}



\begin{abstract}
The red hypergiant VY CMa and the more typical red supergiant Betelgeuse provide clear observational
evidence for discrete, directed gaseous outflows in their optical and infrared imaging, spectra, and light
curves. In the very luminous VY CMa, mass loss estimates from the infrared bright knots and clumps,
not only dominate its measured overall mass loss, but explain it. In the lower luminosity Betelgeuse, similar mass estimates
of its circumstellar condensations show that they contribute significantly to its measured mass loss rate.
We present new measurements for both stars and discuss additional evidence for gaseous ejections in other red
supergiants. Gaseous outflows are   the dominant mechanism for the most luminous RSGs and an important contributor to the more typical red supergiants like Betelgeuse. We conclude that gaseous outflows, related to magnetic fields and surface activity, comparable to coronal mass ejections, are a major
contributor to mass loss from red supergiants and the missing component in discussions of their mass loss mechanism.
\end{abstract}


\section{Introduction}

Mass loss from red supergiants has been well known for decades, but the mechanism 
is not understood.  The leading processes have included radiation
 pressure on dust grains, pulsation, and convection.  Pulsation and dust driven 
winds have been successful at explaining the mass loss of AGB stars, but are not
 adequate for the less variable red supergiants (RSGs) with their very extended,
 low density atmospheres. The presence of large-scale asymmetries on the surfaces 
of RSGs such as Betelgeuse and others lends support for convection as an important source. For examples see \citet{Gilli,Monnier,Haubois,Baron}.    
The recent unexpected dimming of Betelgeuse \citep{Guinan19,Guinan20} corresponded with a remarkable obscuration of its southern hemisphere \citep{Montarges} and 
an outflow of material from the star \citep{Dupree} observed in UV
 spectra. \citet{Montarges} demonstrate that the fading was due to dust formation
 related to the gaseous outflow from a convective cell. They estimate a total mass associated with the dusty and gaseous outflow of 7 $\times 10^{-8}$ to 3 $\times 10^{-7} M_{\odot}$; a significant contributor to Betelgeuse's overall mass loss.  

The history of high mass loss events in the luminous, high mass-losing RSG, VY CMa is clearly visible in its HST image with numerous knots, clumps and large extended 
arcs ejected over several hundred years in different directions from separate regions on the star \citep{RMH2005,RMH2007,RMH19,TJJ}. 
  Mass estimates of some of the knots and clumps, based on IR imaging and ALMA measurements, yield minimum masses of $\approx$ 10$^{-2}$ M$_{\odot}$ \citep{Shenoy,Gordon,OGorman}.  Comparison with its historic light curve from 1800 to the present showed several knots with ejection 
times that correspond with extended periods of variability and deep minima 
lasting years  \citep{RMH2021}.   This similar correspondence of massive outflows with unexpected variability and dust formation with Betelgeuse's brief active period is 
clear, but on a much larger scale in VY CMa. 
Near and mid-infrared imaging of Betelgeuse \citep{Kervella09,Kervella}, 
reveals a circumstellar environment close to the star with arcs and small dusty 
condensations or knots. \citet{Kervella2018} later suggested that convective cells led to the production of dusty knots and molecular plumes in the north polar region based on ALMA observations.  Thus like VY CMa, Betelgeuse, a more typical RSG, also has a history of gaseous outflows from its surface. These two red supergiants, in some sense, may represent the range of episodic mass loss; a typical, relatively young RSG and the more extreme, possibly highly evolved, VY CMa. 

In this short paper, we review the mass estimates for the discrete ejecta in VY CMa and
 add an additional infrared-bright knot close to the star. Following similar procedures, we estimate the masses of the small knots close to Betelgeuse. We describe our measurements
 in the next section. In \S {3}, we discuss the resulting mass loss rates for the 
two stars, and implications for their mass loss mechanism. In \S {4}, we examine evidence from other RSGs for similar outflows. In the final section we conclude that episodic outflows from large active regions is a primary mechanism for red supergiant mass loss.  

\section{Measurements and Mass Estimates of Resolved Ejecta}

Estimates of the mass in some of the arcs and clumps in VY CMa's ejecta based on surface photometry and near-infrared imaging and polarimetry of the SW Clump, 
yield minimum masses of $\approx$ 10$^{-2} M_{\odot}$. ALMA sub-millimeter observations reveal prominent dusty clumps near the star, with the brightest Clump C having a total mass of at least 5 $\times$ 10$^{-2} M_{\odot}$. These measurements are summarized in Table 1.

We have identified an additional  bright knot of emission in its 1$\mu$m HST image  \citep{Smith,RMH2007}, knot W2, 0$\farcs$48 NW of the central star (see Figure 1 in \citet{RMH19}) ). Using the same procedure described in \citet{Shenoy} for their analysis of the SW Clump in 2 --5$\mu$m images of VY CMa, we find that the W2 knot is also optically thick to scattering. With the same dust parameters, we derive a minimum dust mass of $\approx$ 1.7 $\times 10^{-5} M_{\odot}$, compared to $\approx$ 5 $\times 10^{-5}$ for the SW Clump. The difference is primarily due to the smaller angular size of the W2 knot compared to the SW clump. Assuming a gas to dust ratio of 200 \citep{Mauron}, the lower limit to its total mass is  $\approx$ 3.4 $\times 10^{-3} M_{\odot}$. 

The SW clump in VY CMa was also observed in thermal emission in the 10$\mu$m band by \citet{Gordon}.  They found the SW clump was optically thick in emission as well, and derived a minimum dust mass of $\approx$ 5 $\times 10^{-5}$ M$_{\odot}$ , in very good agreement with the results from the observations of scattered light. 

\citet{Kervella} presented  7 -- 20 $\mu$m diffraction limited imaging with VISIR of the dust surrounding Betelgeuse. Combining all of their observations, they identifed  six condensations or knots  within 2\arcsec of the star labeled A -- F. They derived surface brightness fluxes for these features at an effective wavelength of $\lambda$ $\approx$ 13$\mu$m, which we use to estimate the mass in each of the knots in a manner similar to the analysis of thermal emission from the SW Clump in VY CMa. 

Gordon et al.  used the DUSTY code to determine the emission from optically thick dust at the distance of the SW clump from the central star in VY CMa, but this is not necessary for an analysis of the clumps surrounding Betelguese. Adopting the same dust parameters used for VY CMa, we find that, unlike the SW Clump in VY CMa, clumps A-F close to Betelgeuse are all very optically thin. This is easily shown by computing the effective temperature of the dust  using the input dust parameters used in DUSTY from Gordon et al.

The distance to Betelgeuse is uncertain with published distances ranging from 152pc from the revised Hipparcos parallax \citep{Leeuwen} to 220 pc \citep{Harper} based on radio positions and proper motions. The results for the masses of the knots, their distances from the star and ages depend on the adopted distance. In this discussion we show the results for the two most recent results, 168 pc based on asterseismic and hydrodynamical modeling \citep{Joyce} and 220 pc from \citet{Harper}.  The corresponding luminosity of Betelgeuse  is respectively,  1.26 $\times$ 10$^{5}$ L$_{\odot}$  to 2.16 $\times$ 10$^{5}$ L$_{\odot}$ . 

The optical depth is simply

\[\tau  = \frac{{{S_\lambda }}}{{{B_\lambda }(T)}}\]

where S is the observed surface brightness from Figure 11 in \citet{Kervella} 
at an effective wavelength of 13 $\mu$m, and B is the Planck function for the dust temperature at the distance of the relative knot from the star. This temperature is usually somewhat higher than the black-body equilibrium temperature of the dust due to differing absorption efficiencies between the wavelengths of the incident radiation from Betelgeuse and the emitted wavelength (see the discussion and Table 2 in \citet{Gordon} for an example).

Using the same size distribution and optical parameters for the dust from Gordon et al., we can estimate the dust mass column depth in each feature. The effective beam size diameter used by Kervella et al. in their analysis of the knots is 0$\farcs$38, corresponding to 64 AU and 84 AU at  distances  of 168 pc and 220 pc, respectively. Using our estimate of the dust mass column depth and the corresponding surface areas, we computed the range in the dust mass for each feature in Table 2 for the two distances. The range in  computed masses is due to  the  combined uncertainties in the fluxes and the dust model parameters. The range in the total dust mass for the six condensations and the total mass (gas + dust ) is also summarized. We  chose a gas to dust ratio of 200 for comparison with the results for Betelgeuse in \citet{Montarges}.  In the following discussion, we adopt total masses of 2.1 $\times$ M$^{-6} \pm 0.07$ M$_{\odot}$ and 3.6 $\times$ M$^{-6} \pm1.4$ M$_{\odot}$, respectively, for the two distances.   

Based on high spatial resolution visual and near-infrared imaging of Betelgeuse, \citet{Montarges} concluded that the recent dimming of Betelguese can be interpreted as due to a large obscuration of a part of the stellar disk by dust that condensed not far above the photospere. Based on the amount of dimming and the size of the obscuration on the disk, their  rough estimate of the dust mass necessary to explain the dimming was 0.3-1.3 $\times 10^{-9}$ M$_{\odot}$.  The upper range  is  similar to the dust mass estimates we have derived for the six knots surounding Betelgeuse. This suggests that if the recent dimming is due to the formation of dust in a mass-loss event, this new outflow or clump will travel out from Betelgeuse over time and resemble the knots or condensations observed in emission in the 7 -- 20$\mu$m regime.

\begin{deluxetable*}{llll}
\tabletypesize{\footnotesize}
\tablenum{1}
\tablecaption{Mass Estimates for Knots and Clumps in VY CMa}
\tablewidth{0pt}
\tablehead{
\colhead{Knot Id.} & 
\colhead{Dust Mass M$_{\odot}$} & 
\colhead{Total Mass  M$_{\odot}$} & 
\colhead{Reference} \\ 
\colhead{} &
\colhead{}  &
	\colhead{gas/dust 200} &
\colhead{} 
}
\startdata 
	NW Arc  & \nodata      &  $>$ 4  $\times$ 10$^{-3}$  & surface phot \citet{Smith} \\
SW Clump  & $>$ 5 $\times$ 10$^{-5}$ & $>$ 10$^{-2}$   & \citet{Shenoy} \\
	  "  & $>$ 5.4 $\times$ 10$^{-5}$ & $>$ 1.1 $\times$ 10$^{-2}$ & \citet{Gordon} \\   
ALMA Clump C & 2.5 $\times$ 10$^{-4}$ & 5 $\times$ 10$^{-2}$ & \citet{OGorman}    \\
	"   & 1.2 $\times$ 10$^{-3}$  & 0.24 & \citet{Vlemmings} \\ 
W2 knot   & $>$ 1.7 $\times$ 10$^{-5}$ & $>$ 3.4 $\times$ 10$^{-3}$  &  this paper 
\enddata
\end{deluxetable*}

\begin{deluxetable*}{cll}
\tabletypesize{\footnotesize}
	\tablenum{2}
\tablecaption{Mass Estimates for the Knots in Betelgeuse}
\tablewidth{0pt}
\tablehead{
\colhead{Knot Id.} & 
\colhead{Dust Mass M$_{\odot}$} & 
\colhead{Dust Mass  M$_{\odot}$} \\ 
	\colhead{}  &
	\colhead{at 168 pc} &
\colhead{at 220 pc}
}
\startdata 
	A       & 2.0 -- 4.0 $\times$ 10$^{-9}$   & 3.4 -- 6.9 $\times$ 10$^{-9}$ \\  
        B	& 0.9 -- 1.9 $\times$ 10$^{-9}$ & 1.6 -- 3.2  $\times$ 10$^{-9}$    \\
	C       & 1.0 -- 2.0  $\times$ 10$^{-9}$  & 1.7 -- 3.4 $\times$ 10$^{-9}$    \\
	D        &  1.2 -- 2.4 $\times$ 10$^{-9}$  & 2.1 -- 4.1 $\times$ 10$^{-9}$   \\
	E         & 1.4 -- 2.8 $\times$ 10$^{-9}$ & 2.4 -- 4.8 $\times$ 10$^{-9}$   \\
	F        & 0.6 -- 1.2 $\times$ 10$^{-9}$  & 1.0 -- 2.1 $\times$ 10$^{-9}$    \\
	Total Dust Mass &   7.1 -- 14.3 $\times$ 10$^{-9}$  & 1.2 -- 2.4 $\times$ 10$^{-8}$ \\  
	Total Mass (gas {+} dust) &  1.4 -- 2.8 $\times$ 10$^{-6}$ & 2.4 -- 4.8 $\times$ 10$^{-6}$  
\enddata
\end{deluxetable*}

\section{Results -- the Mass Loss Rates} 

The mass estimates in Table 1 for VY CMa are for four different knots or clumps with independent measures. The dusty knots are optically thick in the near and mid-infrared, thus those estimates are lower limits.  The \citet{Vlemmings} estimate for Clump C from 178 GHz continuum observations is obviously high with respect to the others. Deleting it, the average mass lost is about  2 $\times$ 10$^{-2}$ M$_{\odot}$, somewhat greater than the mass of Jupiter. Adopting the average outflow velocity of the inner knots of 27.5 km s$^{-1}$ \citep{RMH2021} the kinetic energy of each ejection is on the order of 10$^{44}$ ergs equivalent to the Sun's radiation in a thousand years.   

Adopting a minimum mass of 10$^{-2}$ M$_{\odot}$ for the gaseous outflows associated with the deep minima in VY CMa's light curve, it shed $\approx$ 7 $\times$ 10$^{-2}$ M$_{\odot}$ during its active period with 7 minima in the early 20th century.  During this 20 year period, from 1925 - 1945, VY CMa's corresponding mass loss rate was thus  $\approx$ 3 $\times$ 10$^{-3}$ M$_{\odot}$ yr$^{-1}$. The corresponding energy expended is a relatively modest one thousandth of the star's total luminosity.   VY CMa's light curve shows three periods with deep minima separated by about 60 years, although the onset and end of the late 19th century minima is uncertain. To get an estimate of of its overall mass loss rate, we assume that over a 100 year period, VY CMa loses mass at the  above rate about 20\% of the time with a quiescent rate of 5 $\times$ 10$^{-6}$  M$_{\odot}$ yr$^{-1}$ \citep{OGorman}. The corresponding  net rate is thus  6 $\times$ 10$^{-4}$ M$_{\odot}$ yr$^{-1}$, comparable to the observed rates of 4 --6  10$^{-4}$ M$_{\odot}$ yr$^{-1}$ from several sources \citep{Shenoy,Danchi}. For VY CMa, its massive gaseous outflows don't just dominate its current mass loss, they explain it. 

The situation is different for Betelgeuse. The total mass in its dusty condensations in Table 2, 2.1 to 3.6 $\times$ 10$^{-6}$ M$_{\odot}$, is more than a thousand times less than in the separate knots in VY CMa. To estimate the mass loss rates from these knots, we need their ages, when they were expelled, which 
depends both on their distance from the star and  their outflow velocity. Velocity measurements from clouds of gas in Betelgeuse's circumstellar environment range from 12 to 18 km s$^{-1}$  \citep{Smith09,Bernat}, and \citet{Dupree} gives 7 km s$^{-1}$ for the recent outflow. 

 These are radial velocities. The total velocity will undoubtedly be higher and yield a lower age.  These age estimates also do not take into account possible projection effects which increase the distance from the star leading to a somewhat higher age. Depending on their size, these two effects tend to counteract each other. Second epoch images are necessary to determine the transverse  
velocity, the direction of the outflows, and   their orientation. Pierre Kervella (private communication) reports that additional VISIR images were obtained during the 2019-2020 dimming which they are currently comparing with the 2010 images.  For this discussion, we adopt an outflow velocity of 10  km s$^{-1}$ and  assume that the condensations are in the plane of the sky. Table 3 summarizes their distances in arcsec and AU  for the two distances and  their estimated ages.

\begin{deluxetable*}{lccccc}
\tabletypesize{\footnotesize}
\tablenum{3}
\tablecaption{Distances and Age Estimates for Knots in Betelgeuse}
\tablewidth{0pt}
\label{tab:Ages}
\tablehead{
\colhead{Knot Id.} & 
\colhead{Distance} & 
	\colhead{Distance (AU)} & 
	\colhead{Age (yrs)} & 
        \colhead{Distance (AU)} & 
	\colhead{Age (yrs)}
	\\ 
\colhead{} &
\colhead{Arcsec}  &
\colhead{ at 168 pc} &
	\colhead{ at 168 pc} & 
	\colhead{ at 220 pc} & 
	\colhead{ at 220 pc} 
}
\startdata 
	A     & 0.88   & 148   & 71  & 194 & 93    \\
	B     & 0.81  &  136  & 65   & 178 & 85   \\
	C     & 1.05  &  176  & 84   & 231 & 111    \\
	D     & 1.48  &  248  & 119  & 325 & 156    \\
	E     & 1.66  &  279  & 134  & 365 & 175    \\
	F   &  1.65   &  277  & 133  & 363 & 174    
\enddata
\end{deluxetable*}

The distances and ages in Table 3 and the spatial distribution of the knots with respect to the star, suggest that they  divide into two groups, ABC and DEF. At the nearer distance, their respective ages are 73 $\pm{8}$ and 129 $\pm{7}$ yrs. With a timescale of 56 $\pm{7}$ yrs, and a total mass of 2.1 $\pm{0.7}$ $\times$  10$^{-6}$ M$_{\odot}$, the mass loss rate from the knots is 3.7 $\pm{1.3}$ $\times$  10$^{-8}$ M$_{\odot}$ yr$^{-1}$.  At the larger distance of 220 pc, their average ages are 96 $\pm{10}$ and 168 $\pm{9}$ yrs. Thus over a timescale of 72 $\pm{9.5}$ yrs, they shed a  mass of  3.6 $\pm{1.2}$ $\times$  10$^{-6}$ M$_{\odot}$ for a comparable rate of  5 $\pm{1.8}$ $\times$  10$^{-8}$ M$_{\odot}$ yr$^{-1}$. The two distances yield essentially the same mass loss rates within their errors. 

The published mass loss rates for Betelgeuse range from a low of 2 $\times$ 10$^{-7}$ M$_{\odot}$ yr$^{-1}$ up to 2 $\times$ 10$^{-6}$ M$_{\odot}$ yr$^{-1}$ \citep{DeBeck,Dolan}. With the lower rate, the gaseous outflows contribute $\approx$ 20\% over these time scales,  but with the higher number the contribution is only a few percent. These numbers are consistent with the \citet{Montarges} estimate of the contribution from the recent outflow to the annual mass loss from Betelgeuse. 

To extend the timescale, we can add the small condensations close to the star, labeled 1,2,3 in \citet{Kervella}. They are at about the same distance from the star, 0$\farcs$5. With the same outflow velocity,  they were all ejected at about the same time, approximately 42 to 55 yrs ago. Assuming the average mass for each  knot, with the additional time span and the increase in total mass,  the resulting mass loss rates  are essentially the same as those quoted above.  

Although examination of Betelgeuse's historic light curve did not reveal any deep minima comparable to the 1990-2020 event \citep{Montarges}, we compared the correspond calendar dates for the outflows with     the AAVSO visual light curve for Betelgeuse. The light curve  from 1900 to the present, shows a broad dip circa 1940 (1938-42) and three  minima in the mid 1940's \footnote{Measured relative to the date of the observations, 2010} with minimum magnitudes of $\sim$ 1.5 mag. Either of which could correspond to the ejection age of knots ABC at the nearer distance. At the larger distance, knots ABC would correspnd to the early 20th century when the observations are sparse.  Knots DEF were apparently ejected before 1900, circa 1880 or 1840. The historic light curve from Baxendell (1840-1884)\footnote{The historic light curves from Herschel, Argelander and Baxendell were kindly provided by Tom Calderwood of the AAVSO} does not show any obvious minima or significant changes in its visual magnitude near these dates, although there are significant gaps in  the observations.  We also find no apparent features in the light curve that would correspond to the innermost condensatons, knots 1,2, 3.  The possible outburst dates however could be uncertain by several years  due to the assumed outflow velocity and the unknown projection.  Furthermore, the outflows may have been out of our direct line of sight or in directions that did obscure a significant fraction of the star.

The large gaseous outflows in VY CMa explain its high mass loss rate, while for the younger and lower mass Betelgeuse, the contribution is less certain with the factor of 10 range in its overall mass loss rate. But assuming that the correct mass loss rate is within the published range, the episodic outflows are also contributing significantly to Betelgeuse's overall mass loss history.

These relatively massive outflows or ejections presumably from the large asymmetries or convective cells observed on the surfaces of RSGs, imply the existence of strong magnetic fields.
 Measurements of circular polarization in SiO, H2O and OH and Zeeeman splitting of  maser emission from the envelope of luminous cool stars have been interpreted as indications of  magnetic fields  in their winds and ejecta. \citet{Vlemmings02} measured the circular polarization of H$_{2}$O maser emission in several luminous, cool stars, including VY CMa for which they derive a field strength of ~200 mG at a distance of ~220 AU from the star. Combining magnetic field strength estimates from OH, H$_{2}$O and SiO maser emission, and extrapolating  the field strength back to the photosphere of the star we estimate a $\approx$ 500 G field assuming a 1/r$^{2}$ Solar-type dependence (Figure 15; \citet{Vlemmings02}). Recently \citet{Shinnaga} directly measured the Zeeman effect in the SiO v = 0, J = 1-0 transition in the envelope of VY CMa, and derive field strengths in the emitting region of anywhere from 10 G (lower limit) to 500 G (upper limit). They conclude that VY CMa must have very strong magnetic field strengths in its mass-loss wind.  Furthermore, continuun images with ALMA \citep{Vlemmings} reveal polarized dust emission from magnetically aligned grains on sub-arcsec scales close to the star consistent with a magnetic field strenth of 1 -- 3 G in the inner wind. 

It is tempting to interpret the large, distinct mass-loss events seen in VY CMa as a scaled up version of large eruptive events (Coronal Mass Ejection, CME) in the Sun.  In their analysis of the global energetics of recent Solar CME events, \citet{Emslie} find that the available magnetic energy is sufficient to power the Coronal Mass Ejections. The largest documented Solar CME was the Carrington event in 1859, with a total kinetic energy of K.E. $\>$ $\approx$ 10$^{33}$ ergs \citep{Cliver}. This event would have required field strengths of B $>$ 200 G in a volume diameter just above the surface of D $\approx$ 0.1R$_{\odot}$. Scaling to VY CMa, the expected kinetic energy of the clumps or knots  is K.E. $\geq$ 10$^{43}$ ergs, and would require a field strength of B $\geq$ 1000 G in a 
volume diameter $\sim$ 1 AU (0.1R$_{star}$) at the surface, comparable to our 
estimates.  

 Comparable  magnetic fields have been measured in the ejecta of other RSGS with maser emission such as VX Sgr, NML Cyg and S Per \citep{Vlemmings02,Vlemmings05} discussed in the next section.  Betelgeuse is not a maser sourse, but \citet{Auriere} and \citet{Mathias} report circular polarization in the photospheric absorption lines due to a longitudinal magnetic fields of $\approx$ 1 G which may be associated with the giant convection cells on its surface. \citet{Tessore} detect similar Zeeman signatures in the RSGs CE Tau and $\mu$ Cep with comparable surface fields also at the one G level. 

We would expect X ray emission to likewise be associated with the magnetic activity and the large outflow. However, VY CMa was not detected in XMM-Newton observations \citep{Montez} with an upper limit to the average surface magnetic field strength fB $\leq$ 2 $\times$ 10$^{-3}$ G where f is the filling factor.  These observations were obtained in 2012, two decades after VY CMa's most recent active period $\sim$ 1985, when it may have been in a state of lower magnetic activity. Interestingly, Betelgeuse's outflow and dimming was also not measured in X rays \citep{Kashyap} with the Chandra X-ray telescope. In both cases the X ray observations were obtained after the active period.

\section{Evidence for Gaseous Outflows from other RSGs}

Without direct observation of a gaseous outflow as in Betelgeuse or a history of episodic mass loss events as observed in VY CMa, it will be difficult to confidently identify  outflows separate from other variability common to red supergiants. A frequent record of photometry and spectroscopy perhaps accompanied by direct imaging, when possible, is needed. But such a record is lacking for most red supergiants. In this section we briefly discuss a few promising stars with high mass loss rates, dusty environments, and resolved circumstellar ejecta.

The luminous OH/IR stars NML Cyg, S Per and VX Sgr are obvious candidates. Like VY CMa, they have measured magnetic fields in their ejecta \citep{Vlemmings02} from the polarization of their water masers that when extrapolated back to their surfaces also imply surface fields of a few hundred G.  The highly variable, high luminosity VX Sgr has a measured dipole field
from the polarization of the H$_{2}$O masers implying a surface field on the order of 1000 G \citep{Vlemmings05}. All three of these high luminosity RSGs have extended circumstellar ejecta in their visual and red HST images \citep{Schuster}, and measured mass loss rates  from 5 $\times$ 10$^{-5}$ to $\geq$ 10$^{-4}$ M$_{\odot}$ yr$^{-1}$ \citep{Shenoy16,Gordon18}.  
 
In many respects NML Cyg most closely resembles VY CMa. 
\citet{Singh} and \citet{Andrews} have recently identified two gaseous outflows in its molecular 
emission spectrum at mm wavelengths. The outflows are  blue and red-shifted relative to the star's sytemic velocity and appear to be aligned with the tips of its small bean or half moon shaped image \citep{Schuster}. They very likely trace sporadic mass loss events.  In addition, 17 different molecules including carbon species identified in its 1mm spectrum are in common with VY CMa. NML Cyg's peculiar morphology is attibuted to its environment. It is embedded in the Cygnus X super bubble and subject to the intense UV radiation from the numerous hot stars in nearby  Cyg OB2 \citep{Schuster,Schuster09}. Otherwise we suspect that NML Cyg would be accompanied by complex circumstellar ejecta similar to VY CMa.  

It is not known if S Per and VX Sgr have experienced episodic outflows. Both show extended ejecta but without obvious substructure \citep{Schuster}, although
 recent 
 3 to 13$\mu$m images of VX Sgr \citep{Chiavassa} reveal a complex surface morphology with possible surface asymmetries. Another clue is anomalous large variations or deep minima in their light curves similar to VY CMa in the early 20th century or smaller episodes as in Betelgeuse.  For example, S Per  is a known semi-regular variable typically with visual light variability of at most $\pm$ 1 mag and a  primary period of about 800 days \citep{Kiss}. Its light curve from $\sim$ 1900 to 2020, shows a few isolated deep minima, but the extended period of about 10 years (1990-2000), with deep minima of about four magnitudes,  is exceptional and is reminiscent of VY CMa's variability in the early 20th century. Afterwards, S Per was about a half magnitude fainter, suggesting that this period of increased instability was followed by dust formation. It is tempting to speculate that, as in VY CMa, S Per had a period of increased mass loss possibly due to surface activity and outflows, although enhanced pulsation is possible.  

VX Sgr is a semi-regular variable that at times behaves like a fundamental mode pulsator (a Mira variable) with deep minima, fading by 5 to 6 magnitudes. 
Although some authors suggest VX Sgr may be a ``super AGB'' star \citep{tabe21} based on its Mira-like variability, its fundamental mode pulsation does not rule out a core burning supergiant \citep{heger97}. Current distance estimates\footnote{Gaia parallaxes range from 0.787mas $\pm{0.238}$ DR2 to 0.046mas $\pm{0.187}$ DR3} ranging from 1.5 to 1.7 kpc due to possible association with SgrOB1 and the Sgr spiral arm \citep{RMH75} to 1.3 to 1.4 kpc \citep{tabe21} indicate a luminosity well above the AGB limit, even taking into account the possibility that a few AGB stars occasionally having luminosities above the AGB limit \citep{garc09}. In this paper we treat VX Sgr as a red supergiant.

Several other, high luminosity red supergiants are high mass losing, dusty and known maser sources, such as AH Sco and MY Cep, and are candidates for active surfaces and gaseous outflows. For example, surface structure has been reported on AH Sco \citep{Witt} and an atmospheric analysis reveals extended molecular layers that need to be elevated above the predictions of the models \citep{Arroyo13}.

More typical, less luminous RSGs, like Betelgeuse, with smaller and shorter
 mass loss episodes or gaseous outflows would be harder or less likely to be identified unless the star was being frequently monitored.  However high resolution imaging and spectroscopy of  RSGs is confirming large surface irregularities in more stars. Furthermore, clumpy winds observed in  RSGs ranging from less luminous members like $\alpha$ Sco \citep{Ohnaka} to  $\mu$ Cep \citep{Montarges2}, one of the visually brightest red supergiants, may be the consequence of surface activity not manifested in massive outflows.

\section{The Mass-Loss Mechanism for Red Supergiants}

There are problems and complications with the three leading mass loss mechanisms for the red supergiants;  radiation pressure on dust grains, pulsation, and 
convection. Dust driven mass loss which works well for the AGB stars \citep{vanloon,Hofner} would initially seem the most likely for the dusty red supergiants. But  the radiation dependent models depend on a direct relation between the mass loss rate and the inrinsic luminosity of the star which is  not confimed by observations \citep{Mauron}. Although there is a clear \.{M}  dependence on luminosity, radiation pressure on grains  does not explain the scatter in the observations. In our recent discussion of mass loss from red supergiants \citep{RMH20}, we also noted the scatter and recommended that the relation is better represented by a broad band or curve than a tight linear relation.

Most red supergiants are known variables, so it is reasonable to suspect that radial pulsation would play a role by expanding the outer layers and depositing material at radii where dust can condense similar to the  highly variable AGB 
stars. However, it is less efficient for the already very extended low density atmospheres of the RSGs.  Given the prevalence of convective cells now observed on several RSGs, a combination of pulsation and convective activity may explain 
the mass loss from the red supergiants.  For example, the recent gaseous outflow from Betelgeuse and dimming apparently corresponded  with the
maximum photospheric outflow velocity during its 400 day pulsation cycle \citep{Dupree,Montarges,Granzer}.

In-depth studies of the atmospheres of two luminous RSGs, AH Sco 
\citep{Arroyo13,Arroyo15} and V602 Car (HD 97671) \citep{Climent} however show that pulsation and convection alone cannot explain the elevation of material to the molecular and dust formation zones. Note that AH Sco is known source of maser emission and near-infrared imaging of V602 Car reveals large convective regions  in its extended atmosphere. 

An additional mass loss mechanism is required. We suggest that episodic high mass loss outflows very likely driven by large scale convection together with magnetic fields comparable to CMEs is the missing component.  
We have shown here that these gaseous outflows expain the very high mass loss rate in VY CMa and contribute to the mass loss from the lower luminosity Betelgeuse.  

Admittedly, we only have 
two confirmed examples, three with NML Cyg, but we emphasize the similarity of VY CMa to the other RSGs with remarkably high mass loss rates. In addition to  mass loss rates $\sim$ 10$^{-4}$  M$_{\odot}$ yr$^{-1}$, many of the most luminous red supergiants have several shared characteristics including late spectral types (M4 --M7), circumstellar ejecta, maser emission, plus measured magnetic fields in several.  
Outflow events in the more typical, less luminous RSGs may be more difficult to detect and contribute less to the total mass loss rate. Thus the presence or lack of outflows may explain the observed spread in the \.{M} -- luminosity relation. Furthermore, if we remove the known maser sources and high \.{M} stars from Figure 12C in \citet{RMH20}, the relation is more linear. This may be circumstantial, but is suggestive that an additional mass loss method is important for these RSGs.  

We therefore suggest that magnetic fields and surface activity resulting in massive gaseous outflows explains the high mass loss rates for these exceptional red supergiants. Gaseous outflows are  the dominant mechanism for the most luminous RSGs and a contributor to the mass loss of the more typical red supergiants like Betelgeuse.

\acknowledgments
This work was supported by NASA through grant GO-15076 (P.I. R. Humphreys) from the Space Telescope Science Institute. We also acknowledge use of the AAVSO visual data light curve for Betelgeuse from 1900 to the present, and thank Tom Calderwood, via private communication to Andrea Dupree, for the early data from Herschel, Argelander and Baxendell. We also  thank Nicolas Mauron and Pierre Kervella for useful comments. 

\vspace{2mm}
\facilities{HST(STIS), AAVSO }

\end{document}